\documentclass[11pt]{article}
 \newcommand{\be}[1]{\begin{equation}\label{#1}}
    \newcommand{\ba}[1]{\begin{eqnarray}\label{#1}}

    \newcommand{\ds}{{\delta\sigma}}
    
    \newcommand{\pa}[1]{\left(#1\right)}
    \newcommand{\paq}[1]{\left[#1\right]}

    \def\ee{\end{equation}}
    \def\ea{\end{eqnarray}}

    \def\s{\sigma}
    \def\ds{\dot\sigma}
    \def\dds{\ddot\sigma}
\usepackage{amssymb,amsmath}
\begin{document}

\title{\textbf{Integrable cosmological models\\ with non-minimally coupled scalar fields}}
\author{\textbf{A.Yu.~Kamenshchik}$^{1,2}$\footnote{Alexander.Kamenshchik@bo.infn.it}, \ \textbf{E.O.~Pozdeeva}$^{3}$\footnote{pozdeeva@www-hep.sinp.msu.ru}, \ \textbf{A.~Tronconi}$^{1}$\footnote{Alessandro.Tronconi@bo.infn.it}, \ \textbf{G.~Venturi}$^{1}$\footnote{Giovanni.Venturi@bo.infn.it}, \ \textbf{S.Yu.~Vernov}$^{3}$\footnote{svernov@theory.sinp.msu.ru}\vspace*{3mm} \\
\small $^1$Dipartimento di Fisica e Astronomia and INFN,\\ \small Via Irnerio 46, 40126 Bologna,
Italy,\\
\small $^2$L.D. Landau Institute for Theoretical Physics of the Russian
Academy of Sciences,\\
\small Kosygin str. 2, 119334 Moscow, Russia,\\
\small $^3$Skobeltsyn Institute of Nuclear Physics, Lomonosov Moscow State University,\\
\small Leninskie Gory 1, 119991, Moscow, Russia}

\date{ \ }
\maketitle

\begin{abstract}
We obtain general solutions for some flat Friedmann universes filled with a scalar field in induced
gravity models and  models including the Hilbert-Einstein curvature term plus  a scalar field conformally coupled to gravity.
As is well known, these models are connected to minimally coupled models through the combination of a conformal transformation and a transformation of the scalar field.
 The explicit forms of the self-interaction potentials for six exactly solvable models are presented here.
We obtain the general solution for one of the integrable models, namely, the induced gravity model with a power-law potential for the self-interaction of the scalar field. We argue that although being mathematically in a one-to-one correspondence  with the solutions in the minimally coupled models, the solutions in the corresponding non-minimally coupled models are physically different. This is because the cosmological evolutions seen by an internal observer connected with the cosmic time can be quite different.
The study of a few induced gravity models with particular potentials gives us an explicit example of such a difference.
\end{abstract}


\section{Introduction}
Exact solutions to the Einstein equations play an important role in cosmology.
Cosmological models with  scalar fields have acquired a great popularity during the last decades due to the development of inflationary cosmology \cite{inflation} and to the appearance of
different quintessence models of dark energy \cite{dark}, which should explain the phenomenon of  cosmic acceleration \cite{cosmic}. The number of integrable cosmological models based on scalar fields is rather limited. First of all, let us mention the spatially flat Friedmann model with a minimally coupled scalar field and a self-interaction exponential potential. For such a  model  the general exact solution was studied in detail in papers \cite{gen-exp}. A particular
power-law expansion solution for this model was found earlier and used in many contexts~\cite{exp-part}. Another interesting class of models with a general exact solution is that
wherein the potential is a combination of exponential functions with special values of the exponents \cite{hyper}. In a recent paper~\cite{Fre}, the general classification of integrable cosmological models based on scalar fields was suggested and studied in great detail (see also~\cite{Fre2}).

 Integrable models are of especial importance for cosmology. For  cosmological models
with minimally coupled scalar fields   some procedures for the reconstruction of potentials were developed. This allows one to obtain models with particular exact solutions~\cite{Onefield}. These procedures have been generalized to two-field models~\cite{Two-fields} and models with scalar fields non-minimally coupled to gravity~\cite{we-ind-ex, we-ind-ex1}. For example, some particular exact solutions, for the models where the scalar field was non-minimally coupled to gravity were found in~\cite{we-ind-ex,we-ind-ex1}. At the same time, to the best of our knowledge there is no generic procedure to obtain integrable cosmological models with scalar fields. The goal of this paper is to show how the knowledge of an integrable system with a minimally coupled scalar field allows one to obtain the integrable system with a non-minimally coupled one and  vice versa.
We illustrate the correspondence between the integrable cosmological models with minimally and non-minimally coupled scalar fields.
We construct the explicit form of the self-interaction potentials for six integrable induced gravity models and six integrable models with
the Hilbert--Einstein curvature term and a scalar field conformally coupled to gravity. We also explicitly obtain the general solutions for induced gravity models with power-law potentials.

Models with a scalar field  non-minimally coupled to gravity  attract a growing interest nowadays. Induced gravity models, wherein the curvature arises as a quantum effect~\cite{Sakharov} have found diverse applications in cosmology~\cite{induced}. Models, where both the Hilbert--Einstein term and the  scalar field squared multiplied by the scalar curvature are present, have been applied to inflation~\cite{nonmin-inf} and to   quantum cosmology~\cite{nonmin-quant}. Recently, a model where the role of the inflaton is played by the Higgs field non-minimally coupled to gravity has acquired some popularity~\cite{Higgs}.
The recent discovery of the Higgs particle \cite{discovery} makes this model especially attractive.

It is known that a gravitational model with a non-minimal coupling can be transformed into a model with a minimal coupling, by using a properly chosen conformal transformation of the metric, combined with some parametric transformation of the scalar field. Such a procedure is usually called the transition from the Jordan frame to the Einstein frame.
For the first time this transformation was used in the seminal paper by Wagoner \cite{Wagoner}.
For the more general case of $F(\phi,R)$ gravity, where $\phi$ is a scalar field, such a transformation was presented in~\cite{Maeda}.
The Wagoner conformal mapping is used to get both cosmological and spherically symmetric solutions. For example,  on using Wagoner's conformal mapping spherically symmetric solutions with and without an electric charge were obtained by Bronnikov in \cite{Bronnikov}.

There is some discussion in the literature about the equivalence or non-equivalence of physics in these two frames \cite{Dicke,Sasaki,Sasaki1,frames}.
The dominant opinion is that for non-conformal matter these frames are not equivalent.
The question about  which of the frames is physical is discussed in detail in such papers as \cite{Faraoni} and \cite{Bron-Meln}.
 In particular, in paper \cite{Bron-Meln} it is claimed, that the word ``physical'' applied to conformal frames can have two meanings: either the frame in which a fundamental underlying theory is formulated, or the one corresponding to the observational picture, and these two do not necessarily coincide.

 Here we wish to draw attention to another aspect of the interrelation between the physics in the  Jordan and Einstein frames.  We consider some explicitly integrable cosmological models for flat Friedmann universes filled with a minimally coupled scalar field, taken from paper~\cite{Fre} and, on using the Einstein--Jordan frame transition, we construct their counterparts with induced gravity models and with models having a conformally coupled scalar field plus the Hilbert--Einstein term.
We emphasize that the physical cosmological evolutions are those seen by an observer using the cosmic (synchronous) time, which is different in different frames. Thus, evolutions in the Einstein and Jordan frames, connected by a conformal transformation and by the reparametrization of the scalar field can be qualitatively different. We construct an explicit example of such a difference. More precisely, we consider a de Sitter expansion in induced gravity with a squared scalar field self-interaction potential~\cite{we-ind-ex,Star-unpub} and show that its counterpart is the well-known particular power-law solution~\cite{exp-part} in a minimally coupled model with an exponential potential. We think that while the general solutions for some non-minimally coupled models can be obtained from their minimally coupled counterparts, the study of their behaviour is of interest because it can be physically different.

The structure of the paper is as follows: in Sect. 2 we present the general formulae,
representing the relations between the general exact explicitly integrable cosmological
solutions in  models with minimal and non-minimal coupling of a scalar field with gravity;
in Sect. 3 we construct the potentials for exactly solvable models with a non-minimal coupling
starting from the minimally coupled models studied in \cite{Fre}; in Sect. 4 we find the general solution for induced gravity model with power-law potential; in Sect. 5
we analyze the example with a de Sitter evolution, while the last Section contains some
concluding remarks.
\section{Relations between general exact solutions in the models with minimally coupled and non-minimally coupled scalar fields}
Let us consider an action
\begin{equation}
S =\int d^4x\sqrt{-g}\left[U(\sigma)R - \frac12g^{\mu\nu}\sigma_{,\mu}\sigma_{,\nu}+V(\sigma)\right].
\label{action}
\end{equation}
In  a flat Friedmann space-time with the interval
\begin{equation}
ds^2 = N^2(\tau)d\tau^2 -a^2(\tau)\vec{dl}^2,
\label{Fried}
\end{equation}
where $a(\tau)$ is the cosmological radius and $N(\tau)$ is the lapse function, the Lagrangian corresponding to this action is~\cite{we-ind-ex,KKhT}
\begin{equation}
L=\frac{6\dot{a}^2aU}{N}+\frac{6 a^2\dot{a}\dot{\sigma}U'}{N}-\frac{a^3\dot{\sigma}^2}{2N}+N Va^3,
\label{Lagrange}
\end{equation}
where a ``dot'' means a derivative with respect to time and a ``prime'' means a derivative with respect to $\sigma$.
The variation of this Lagrangian with respect to the lapse function gives the first Friedmann equation:
\begin{equation}
\frac{6U\dot{a}^2}{a^2}+\frac{6U'\dot{a}\dot{\sigma}}{a}=\frac12\dot{\sigma}^2+N^2V.
\label{Fried1}
\end{equation}

The variation with respect to $a$ gives the second Friedmann equation
\begin{equation}
 \frac{4U\ddot{a}}{a}+\frac{2U\dot{a}^2}{a^2}+\frac{4U'\dot{a}\dot{\sigma}}{a}-\frac{4U\dot{a}\dot{N}}{aN}+2U''\dot{\sigma}^2
+2U'\ddot{\sigma}-\frac{2U'\dot{\sigma}\dot{N}}{N} = -\frac12\dot{\sigma}^2+N^2V.
\label{Fried2}
\end{equation}

The variation with respect to  $\sigma$ gives the Klein--Gordon equation:
\begin{equation}
\ddot{\sigma}+\left(3\frac{\dot{a}}{a}-\frac{\dot{N}}{N}\right)\dot{\sigma} -6U'\left[\frac{\ddot{a}}{a}
+\frac{\dot{a}^2}{a^2}\right]+6\frac{\dot{a}\dot{N}}{aN}U'+N^2V' = 0.
\label{KG}
\end{equation}

In what follows it will be convenient to use a certain linear combination of
Eqs.~(\ref{Fried1}) and (\ref{Fried2}):
\begin{equation}
  4U\left[\frac{\ddot{a}}{a}+\frac{\dot{a}^2}{a^2}\right]={}-6U'\dot{\sigma}\frac{\dot{a}}{a}
+4U\frac{\dot{a}\dot{N}}{aN}-2U''\dot{\sigma}^2-2U'\ddot{\sigma}+
2U'\dot{\sigma}\frac{\dot{N}}{N}-\frac{1}{3}\dot{\sigma}^2+\frac{4}{3}N^2V.
\label{Fried12}
\end{equation}

Using (\ref{Fried12}), we write (\ref{KG}) as follows:
\begin{equation}
\label{KGm}
 \left[\ddot{\sigma}+\left(3\frac{\dot{a}}{a}-\frac{\dot{N}}{N}\right)\dot{\sigma}\right]\left[
1+3\frac{{U'}^2}{U}\right]+\frac{U'}{2U}\dot{\sigma}^2\left[1+6U''\right]
+N^2\left[V'-2\frac{U'}{U}V\right]= 0.
\end{equation}

Let us make the conformal transformation of the metric
\begin{equation}
g_{\mu\nu} = \frac{U_0}{U}\tilde{g}_{\mu\nu},
\label{conf}
\end{equation}
where $U_0$ is a constant.
We also introduce a new scalar field $\phi$ such that
\begin{equation}
\frac{d\phi}{d\sigma} = \frac{\sqrt{U_0(U+3U'^2)}}{U}
\quad\Rightarrow\quad
\phi = \int \frac{\sqrt{U_0(U+3U'^2)}}{U} d\sigma.
\label{scal1}
\end{equation}
In this case the action (\ref{action}) becomes the action for a minimally coupled scalar field:
\begin{equation}
S =\int d^4x\sqrt{-\tilde{g}}\left[U_0R(\tilde{g}) - \frac12\tilde{g}^{\mu\nu}\phi_{,\mu}\phi_{,\nu}+W(\phi)\right] ,
\label{action1}
\end{equation}
where
\begin{equation}
W(\phi) = \frac{U_0^2 V(\sigma(\phi))}{U^2(\sigma(\phi))}.
\label{poten}
\end{equation}
Let us emphasize that the formulae (\ref{conf})--(\ref{poten}) are valid for any metric. We  now present their particular forms for the spatially flat Friedmann metric with a parametric time.

The Friedmann metric (\ref{Fried}) becomes
$ds^2 = \tilde{N}^2d\tau^2 - \tilde{a}^2\vec{dl}^2$,
where the new lapse function and the new cosmological radius are defined as
\begin{equation}
\tilde{N} = \sqrt{\frac{U}{U_0}}N,\qquad \tilde{a} =  \sqrt{\frac{U}{U_0}}a.
\label{Na}
\end{equation}
Equations (\ref{Fried1})--(\ref{KG}) become
\begin{equation}
6U_0\tilde{h}^2=\frac12\dot{\phi}^2+\tilde{N}^2W,
\label{Fried10}
\end{equation}
\begin{equation}
4U_0\dot{\tilde{h}}+6U_0\tilde{h}^2-4U_0\tilde{h}\frac{\dot{\tilde{N}}}{\tilde{N}}
= -\frac12\dot{\phi}^2+\tilde{N}^2W,
\label{Fried20}
\end{equation}
\begin{equation}
\ddot{\phi}+\left(3\tilde{h}-\frac{\dot{\tilde{N}}}{\tilde{N}}\right)\dot{\phi}
+\tilde{N}^2W_{,\phi} = 0,
\label{KG0}
\end{equation}
where $\tilde{h} \equiv \dot{\tilde{a}}/{\tilde{a}}$.

Let us suppose that for some potential $W$ we know the general exact solution of the system of equations (\ref{Fried10})--(\ref{KG0}): $\phi(\tau)$, \ $\tilde{a}(\tau)$, \ $\tilde{N}(\tau)$.
We also suppose  that the function~$\sigma(\phi)$ is known explicitly.

In this case, we can  also find the general solution of the system of equations (\ref{Fried1})--(\ref{KG}) with the potential
\begin{equation}
V(\sigma) = \frac{U^2(\sigma)W(\phi(\sigma))}{U_0^2}\,,
\label{poten1}
\end{equation}
which is given by
$\sigma(\phi(\tau))$,\ $a(\tau) = \sqrt{\frac{U_0}{U(\sigma(\phi(\tau)))}}\tilde{a}(\tau)$,\ $N(\tau) = \sqrt{\frac{U_0}{U(\sigma(\phi(\tau))}}\tilde{N}(\tau)$.

Thus, if we choose
\begin{equation}
U(\sigma) = \frac\gamma2  \sigma^2,
\label{induced}
\end{equation}
i.e. the induced gravity case, we have from Eq. (\ref{scal1})
\begin{equation}
 \phi = \sqrt{\frac{2U_0(1+6\gamma)}{\gamma}}\ln \left[\frac{\sigma}{\sigma_0}\right]
\qquad \mbox{and, inversely, } \qquad
\sigma = \sigma_0 e^{\sqrt{\frac{\gamma}{2U_0(1+6\gamma)}}\phi}.
\label{connection1}
\end{equation}
In this case we can construct analogues of all the solutions known for the case of a minimally coupled scalar field. We assume that $\gamma\neq -1/6$ because for the case of  conformal coupling, $\gamma= -1/6$,  nontrivial solutions can exist only for the potential $V=V_0\sigma^4$~\cite{ABGV}. Another interesting choice is
\begin{equation}
U(\sigma) = U_0 - \frac{\sigma^2}{12},
\label{conf-coupl}
\end{equation}
i.e.  the case when the coupling is conformal and a nonzero Einstein--Hilbert term is also present\footnote{The models with a such $U(\sigma)$ have been considered and connected with $f(R)$-gravity models in~\cite{BNOS}.}. In this case
\begin{equation}
\phi = \sqrt{3U_0}\ln \left[\frac{\sqrt{12U_0}+\sigma}{\sqrt{12U_0}-\sigma}\right]
\quad\mbox{and} \quad
\sigma = \sqrt{12U_0}\tanh\left[ \frac{\phi}{\sqrt{12U_0}}\right].
\end{equation}

\section{Integrable Models}

In this paper, we only consider  those integrable cases, mentioned in \cite{Fre}, for which the explicit solutions for $\tilde{N}(\tau)$, $\tilde{a}(\tau)$ and $\phi(\tau)$ can be found at least in quadratures. To make the comparison with the models analyzed in paper \cite{Fre} more convenient, we choose $U_0 = 1/4$.

The first of them is the exponential potential. The general solution for such a model is known and was studied in detail in the literature~\cite{gen-exp}.
Let us consider the scalar field with the potential
\begin{equation}
W = W_0e^{2\sqrt{3}\lambda\phi},
\label{pot-exp}
\end{equation}
where $\lambda$ is  an arbitrary real number, but $\lambda \neq \pm1$.

The corresponding potential in the induced gravity model is
\begin{equation}
V(\sigma) = 4W_0 \gamma^2\sigma^4\left(\frac{\sigma}{\sigma_0}\right)^{\lambda\sqrt{\frac{6(1+6\gamma)}{\gamma}}}=
4W_0 \gamma^2\sigma^4\left(\frac{\sigma}{\sigma_0}\right)^{6\lambda\Gamma}.
\label{exp-ind}
\end{equation}
where $\Gamma\equiv \sqrt{\frac{1+6\gamma}{6\gamma}}$. We can set $\sigma_0=1$ without loss of    generality, because $W_0$ is an arbitrary constant. Note that if $\Gamma$ is real, then $\Gamma>0$ because
$\gamma\neq -1/6$.

For the model with the conformally coupled scalar field we have the potential
\begin{equation*}
  {\cal V}(\sigma) = W_0\left[1-\frac{\sigma^2}{3}\right]^2\left(\frac{\sqrt{3}+\sigma}{\sqrt{3}-\sigma}\right)^{3\lambda}\!=
W_0\Theta\Upsilon^{3\lambda},\quad \Theta\equiv\left[1-\frac{\sigma^2}{3}\right]^2, \ \Upsilon\equiv\frac{\sqrt{3}+\sigma}{\sqrt{3}-\sigma}.
\end{equation*}
\begin{table}[h]
\begin{center}
\caption{Families of integrable potentials in minimally and non-minimally coupled  models}
\begin{tabular}{|l| c| c|c|}
\hline
  & $W$ (minimal coupling) & $V$ (induced gravity) & {$\cal V$} (conformal coupling) \\[2.7mm]
  \hline
$\!1\!$ & $c_0{e^{2\sqrt{3}\lambda\phi}}^{\vphantom{L}}$ & $\tilde{c}_0 \sigma^{4+6\lambda\Gamma^{\vphantom{L}}}$
& $c_0\Theta\Upsilon^{3\lambda}$ \\[2.7mm]
\hline
$\!2\!$ & $c_0+c_1e^{\sqrt{3}\phi}+c_2e^{-\sqrt{3}\phi}\!$ & $\!
\tilde{c}_0\sigma^4+
\tilde{c}_1\sigma^{4+3\Gamma}
+\tilde{c}_2{\sigma^{4-3\Gamma}}^{\vphantom{L}}\!$&
$\!\Theta\left[c_0+c_1\Upsilon^\frac{3}{2}+c_2\Upsilon^{-\frac{3}{2}}\right]^{\vphantom{L}}$  \\[2.7mm]
\hline
$\!3\!$ & $c_1e^{2\sqrt{3}\lambda\phi}+c_2e^{\sqrt{3}(\lambda+1)\phi}$ & $\tilde{c}_1\sigma^{4+6\lambda\Gamma}
+\tilde{c}_2{\sigma^{4+3(\lambda+1)\Gamma}}^{\vphantom{L}}$ &
$\Theta\left[c_1\Upsilon^{3\lambda}+c_2\Upsilon^{\frac{3}{2}(\lambda+1)}\right]^{\vphantom{L}}$\\[2.7mm]
\hline
$\!4\!$ & $c_1e^{2\sqrt{3}\phi}+c_2 $& $\sigma^4\left[\tilde{c}_1\sigma^{6\Gamma}+\tilde{c}_2\right]^{\vphantom{L}}$
& $\Theta\left[c_1\Upsilon^{3}+c_2\right]$ \\[2.7mm]
\hline
$\!5\!$ & $c_0\phi e^{2\sqrt{3}\phi}$& $\sqrt{3}\Gamma\tilde{c}_0\sigma^{4+6\Gamma^{\vphantom{L}}}
\ln\left[\frac{\sigma}{\sigma_0}\right]$ & $\frac{\sqrt{3}}{2}c_0\Theta\Upsilon^{3}
\ln\left(\Upsilon\right)$\\[2.7mm]
\hline
$\!6\!$ & $c_1e^{2\sqrt{3}\lambda\phi}+c_2e^{\frac{2\sqrt{3}}{\lambda}\phi}$ &
$\sigma^4\left[\tilde{c}_1\sigma^{6\lambda\Gamma}
+\tilde{c}_2\sigma^{6\frac{\Gamma}{\lambda}}\right]^{\vphantom{L}}$&
$\Theta\left[c_1\Upsilon^{3\lambda}+c_2\Upsilon^{\frac{3}{\lambda}}\right]$\\[2.7mm]
\hline
\end{tabular}
\label{IntegrablePotentials}
\end{center}
\end{table}

In Table {\ref{IntegrablePotentials}} we present the list of the potentials, for which the corresponding cosmological models are integrable. The constants $c_0$, $c_1$, and $c_2$ are arbitrary, $\tilde{c_i}=4\gamma^2 c_i$. The parameter $\lambda$ is an arbitrary number, but $\lambda\neq \pm 1, \lambda\neq 0$.
\begin{table}[h]
\begin{center}
\caption{Lapse functions for integrable cases}
\begin{tabular}{|l|c|c|c|}
 \hline
 &${\tilde{N}}^{\vphantom{L}}$ (minimal coupling) & $ N$  (induced gravity) & ${\cal N}$  (conformal coupling) \\[2.7mm]
\hline
 1 &$ \frac{\sqrt{6}}{\sqrt{c_0}} e^{-\sqrt{3}\lambda\phi}$ &$ \frac{\sqrt{3}}{\sqrt{\gamma c_0}}\sigma^{-3\lambda\Gamma-1}$ & $\sqrt{\frac{18}{c_0(3-\sigma^2)}}^{\vphantom{L}}\Upsilon^{-3\lambda/2} $\\[2.7mm]
\hline
  2& $1$ & $\frac{\sqrt{2}}{\sqrt{\gamma}\sigma}$ & $ \sqrt{\frac{3}{3-\sigma^2}}^{\vphantom{L}}$ \\[2.7mm]
\hline
 3& $e^{-\sqrt{3}\lambda \phi}$ & $\frac{1}{\sqrt{2\gamma}}\sigma^{-3\Gamma\lambda-1}$ & $\sqrt{\frac{3}{3-\sigma^2}}^{\vphantom{L}}\Upsilon^{-3\lambda/2}$
   \\[2.7mm]
\hline
 4& $ e^{-\sqrt{3}\phi}$ & $\frac{1}{\sqrt{2\gamma}}\sigma^{-3\Gamma-1}$ & $\sqrt{\frac{3}{3-\sigma^2}}^{\vphantom{L}}\Upsilon^{-3/2}$
    \\[2.7mm]
\hline
 5& $\displaystyle\frac{e^{-2\sqrt{3}\phi}}{\tilde{a}^3}$    & $\frac{9(\Gamma^2-1)^2}{a^3\sigma^4}\left(\frac{\sigma}{\sigma_0}\right)^{-6\Gamma} $  & $\frac{9}{a^3}\frac{(\sqrt{3}-\sigma)}{(\sqrt{3}+\sigma)^5} $
    \\[2.7mm]
\hline
    6 &  $\tilde{a}^3 $ & $\frac{\sigma^2a^3}{3(\Gamma^2-1)} $ & ${\left(1-\frac{\sigma^2}{3}\right)^2}^{\vphantom{L}}a^3$ \\[2.7mm]\hline
\end{tabular}
\label{funcN}
\end{center}
\end{table}

The standard way of obtaining the general solution for an integrable model with a minimally coupled scalar field is to find a suitable lapse function $\tilde N(\tau)$.
For the minimally coupled models with the potentials given in
Table {\ref{IntegrablePotentials}}, the corresponding functions $\tilde N(\tau)$ are given in~\cite{Fre}.
In Table \ref{funcN} we present the lapse functions $\tilde{N}$,  $N$ and $\cal N$ corresponding respectively to the minimal coupling, induced gravity and conformal coupling plus the Hilbert--Einstein term case. The functions  $N$ and $\cal N$ have been calculated using~(\ref{Na}).

In the next section, we shall consider in some detail the model of induced gravity with a power-law potential for the self-interaction of the scalar field.
As was explained above the general solution for this model can be obtained by using the solutions of the model with a minimally coupled scalar field and with an exponential potential.
However, in this case these general solutions can also be obtained directly, because the
knowledge of the solutions for the minimally coupled model can help us to guess what choice of the variables and of the lapse function can be useful for the exact integration of the corresponding equations (\ref{Fried1})--(\ref{KGm}) in the induced gravity model.

\section{Induced gravity model with a power-law potential}
\label{MonomialPotential}
\subsection{Construction of linear equations}
Let us consider the example of an integrable induced gravity model to demonstrate how the results of
the previous section help in obtaining the general solution. In the previous sections we assumed
that the functions~$\sigma(\phi)$ and $\phi(\tau)$ were known explicitly. In this section we shall not use these functions
to obtain the general solution of the induced gravity model.

The first Friedmann equation (\ref{Fried1}) with $U(\sigma)$, defined by (\ref{induced}),
can be cast in the following form
\be{freq2}
\pa{\frac{d}{d\tau}\ln a\s}^{2}-\pa{\frac{d}{d\tau}\ln\sigma^{\Gamma}}^{2}=\frac{VN^2}{3\gamma\sigma^2}.
\ee

We consider the power-law potential
\begin{equation}
\label{plot}
    V=4\gamma^2 c_0 \sigma^{2n},
\end{equation}
where $n=2+3\lambda\Gamma$, corresponds to the integrable Einstein gravity model with an exponential
potential (\ref{pot-exp}).

From Table~\ref{funcN} we learn that the suitable choice of the lapse function is
\begin{equation}
N=\frac{\sqrt{3}}{\sqrt{\gamma c_0}}\sigma^{1-n}.
\end{equation}

On introducing new variables $u$ and $v$ as
\be{uvdef}
a\s\equiv e^{{u+v}}\;,\qquad \sigma^{\Gamma}\equiv e^{{u-v}},
\ee
we see that Eq.~(\ref{freq2}) takes the form:
\be{freq4}
\dot u\dot v=\frac{VN^2}{12\gamma\sigma^2}=1.
\ee

Equation (\ref{uvdef}) implies  that
\be{ds/s}
\frac{\ds}{\s}=\frac{1}{\Gamma}\pa{\dot{u}-\dot{v}},
\qquad
\frac{\dds}{\s}=\frac{1}{\Gamma^{2}}\paq{\Gamma\pa{\ddot{u}-\ddot{v}}+\pa{\dot{u}-\dot{v}}^{2}},
\ee
\be{H}
 \frac{\dot{N}}{N}=\frac{1-n}{\Gamma}\pa{\dot{u}-\dot{v}},
\qquad
\frac{\dot{a}}{a}=\frac{1}{\Gamma}\left[(\Gamma-1)\dot{u}+(\Gamma+1)\dot{v}\right],
\ee

Let us now consider Eq.~(\ref{KGm}) for the induced gravity model:
\begin{equation}
\label{KGIG}
\left[\ddot{\sigma}+\left(3\frac{\dot{a}}{a}-\frac{\dot{N}}{N}\right)\dot{\sigma}
+\frac{\dot\sigma^2}{\sigma}\right]\left(1+6\gamma\right)
+\left[V'-\frac{4}{\sigma}V\right]N^2=0.
\end{equation}

On substituting the functions $N$ and $V$, we obtain the following equation:
\begin{equation}
\label{KGIGm}
\frac{\ddot{\sigma}}{\sigma}+3\frac{\dot{a}\dot{\sigma}}{a\sigma}
-\frac{\dot{N}
\dot{\sigma}}{N\sigma}+\frac{\dot\sigma^2}{\sigma^2}+\frac{4(n-2)}{\Gamma^2}
= 0
\end{equation}
and on using (\ref{ds/s}) and (\ref{H}), Eq.~(\ref{KGIGm}) can be rewritten in terms of variables $u$ and $v$:
\be{kgeq2}
\Gamma\pa{\ddot{u}-\ddot{v}}+(n-2)(\dot{u}-\dot{v})^2+3\Gamma\left(\dot{u}^2-\dot{v}^2\right)+4(n-2)=0.
\ee

We use Eq.~(\ref{freq4}) and introduce a new variable
\begin{equation}
\label{x}
    x= \dot u\qquad\Rightarrow\qquad \dot v=\frac{1}{x}\,.
\end{equation}

It is easy to check that Eq.~(\ref{kgeq2}) is equivalent to the following Riccati equation
\be{kgeq3}
\dot{x}+\frac{n-2+3\Gamma}{\Gamma}x^{2}+\frac{n-2-3\Gamma}{\Gamma}=0.
\ee

To construct the general solution of Eq. (\ref{kgeq3}) , we make a standard substitution
\be{subRic}
x=\frac{\Gamma\dot{y}}{(n-2+3\Gamma)y}\,,
\ee
leading to  the second order linear differential equation:
\be{kgeq4}
\ddot y+ky=0\,,
\ee
where $k=\frac{\pa{n-2}^{2}}{\Gamma^{2}}-9\equiv l^{2}-9$ and $l=(n-2)/{\Gamma}$. Let us note that all solutions of type $Cy(t)$, where $C$ is a nonzero constant, correspond to one and the same solution $x(t)$. Further,
Eq.~(\ref{kgeq4}) has a solution $y(\tau)\equiv 0$ for any $k$, but $x$ is not defined for this solution.

\subsection{Different types of solutions}
Let us now consider the different types of solutions arising in the model under consideration.
let us note that, $x$ can not be defined at $n=2-3\Gamma$, but in this case Eq.~(\ref{kgeq3}) is linear one we not need to introduce~$y$.
Further, it is simpler to consider constant solutions of Eq.~(\ref{kgeq3}) as well as the special case $k=0$ and $n=2+3\Gamma$
without the introduction of the variable $y$.
We consider  the cases $k>0$,  $k<0$, and the cases for which  it is more suitable to solve Eq.~(\ref{kgeq3}) separately.

\subsubsection{Trigonometric case}

{ \ }

In the case $k=\Omega^{2}>0$, the general solution of (\ref{kgeq4}) is
\be{Trsol}
y=A\cos{(\Omega\tau)}+B\sin{(\Omega\tau)},
\ee
where $A$ and $B$ are arbitrary constants.
We get
\begin{equation}
x=\frac{\Gamma\Omega(B\cos(\Omega\tau)-A\sin(\Omega\tau))}{(n-2+3\Gamma)(A\cos(\Omega\tau)+B\sin(\Omega\tau))},
\end{equation}
Let us note that when $A$ and $B$ are not equal to zero, $x$ is defined by the arbitrary parameter~$A/B$.
\be{Trsolu}
u=u_{0}+\frac{\Gamma}{n-2+3\Gamma}\ln\left(A\cos(\Omega\tau)+B\sin(\Omega\tau)\right),
\ee
\begin{equation}
v=v_0-\frac{n-2+3\Gamma}{\Gamma{\Omega}^2}\ln{(A\sin(\Omega\tau)-B\cos(\Omega\tau))}.
\end{equation}

On using (\ref{uvdef}), we obtain
\begin{equation}
\sigma=\sigma_0[A\sin(\Omega\tau)-B\cos(\Omega\tau)]^{1/(n-2-3\Gamma)}
[A\cos(\Omega\tau)+B\sin(\Omega\tau)]^{1/(n-2+3\Gamma)}\,,
\end{equation}
\begin{equation}
  a=a_0[A\sin(\Omega\tau)-B\cos(\Omega\tau)]^{(\Gamma+1)/(3\Gamma-n+2)}
[A\cos(\Omega\tau)+B\sin(\Omega\tau)]^{(\Gamma-1)/(3\Gamma+n-2)},
\end{equation}
where $\sigma_0=e^{\frac{u_0-v_0}{\Gamma}}$ and $a_0=e^{u_0+v_0}/\sigma_0$.

\subsubsection{Hyperbolic case}

{ \ }

For the case $k={}-\tilde{\Omega}^{2}<0$, $\tilde{\Omega}=\sqrt{9-l^{2}}$, the general solution of (\ref{kgeq4}) is
\be{hsol}
y=c_{+}e^{\tilde{\Omega} \tau}+c_{-}e^{-\tilde{\Omega} \tau}
\ee
leading to
\be{hsolu}
u=u_{0}+\frac{1}{3+l}\ln\pa{c_{+}e^{\tilde{\Omega} \tau}+c_{-}e^{-\tilde{\Omega} \tau}}
\ee
and
\be{hsolv}
v=v_{0}+\frac{1}{3-l}\ln\pa{c_{+}e^{\tilde{\Omega} \tau}-c_{-}e^{-\tilde{\Omega} \tau}},
\ee
where $c_+$ and $c_-$ are arbitrary constants. If $c_+=0$ or $c_-=0$, then $x$ is a constant. In this subsection we consider only
solutions having $c_+c_-\neq 0$, solutions with a constant $x$ are considered in the next subsection.

On inverting the definitions for the functions $u$ and $v$ (\ref{uvdef}), one  finds
\be{hsigma}
\sigma=\sigma_{0}\frac{\left(c_{+}e^{\Omega \tau}+c_{-}e^{-\Omega \tau}\right)^{1/(3\Gamma+n-2)}}{\left(c_{+}e^{\Omega \tau}-c_{-}e^{-\Omega \tau}\right)^{1/(3\Gamma-n+2)}}\,,
\ee
\begin{equation}
\label{ha}
a=a_{0}\pa{c_{+}e^{\Omega \tau}+c_{-}e^{-\Omega \tau}}^{\frac{\Gamma-1}{3\Gamma+n-2}}\pa{c_{+}e^{\Omega \tau}-c_{-}e^{-\Omega \tau}}^{\frac{\Gamma+1}{3\Gamma-n+2}},
\end{equation}
where  $\sigma_0=e^{(u_0-v_0)/\Gamma}$ and $a_0=e^{u_0+v_0}/\s_{0}$.

\subsubsection{Solutions of Eq.~(\ref{kgeq3}) in special cases and constant solutions of this equation.}

{ \ }

Equation~(\ref{kgeq3}) has constant solutions
\begin{equation}
\label{x0}
x=x_0\equiv{}\pm\sqrt{\frac{n-2-3\Gamma}{2-n-3\Gamma}}.
\end{equation}

Let us consider solutions having $x=x_0\neq 0$. For a given $\Gamma$ these solutions are real for
\begin{equation*}
\Gamma>\frac{|n-2|}{3}.
\end{equation*}
For $n=2$ solutions with
$x=x_0$ exist for any  value of $\Gamma$ (obviously, $\Gamma$ ranges in the interval $1 < \Gamma < \infty$ at $\gamma>0$).
For these solutions we obtain
\begin{equation*}
u=x_0(\tau-\tau_1), \qquad v=\frac{1}{x_0}(\tau-\tau_2),
\end{equation*}
where $\tau_1$ and $\tau_2$ are arbitrary constants. Therefore,
\begin{equation}
\label{sigma_a_x0}
a=a_0e^{((\Gamma-1)x_0^2+(\Gamma+1))\tau/(x_0\Gamma)},\qquad \sigma=\sigma_0e^{(x^2_0-1)\tau/(\Gamma x_0)},
\end{equation}
where $\sigma_0=e^{(\tau_2-x_0^2\tau_1)/(\Gamma x_0)}$, \ $a_0=e^{(1-\Gamma)x_0\tau_1/\Gamma-(1+\Gamma)\tau_2/(x_0^2\Gamma)}$.

Let us note that for $n=2$, we get $x_0=1$ and $\sigma(\tau)$ is a constant. In this well-known case of a quartic potential ($n=2$) the solution is known to have a de Sitter attractor in the future and correspondingly a constant scalar field $\sigma(\tau)=\sigma_{f}$. Such an attractor is easily recovered in the large $\tau$ limit from solutions (\ref{hsigma})--(\ref{ha}) when
\be{DSa}
\sigma(\tau)\rightarrow \sigma_f,\qquad \qquad a(\tau)\rightarrow a_{DS}(\tau)\sim e^{\pm 2\tau}.
\ee
It is suitable to study physical properties of solutions in the cosmic time (see the next subsection).

Let us consider the case $n=2-3\Gamma$, in which case Eq.~(\ref{kgeq3}) is linear and has the following general solution:
\begin{equation}
x=6(\tau-\tau_0),
\end{equation}
\begin{equation}
u=3(\tau-\tau_0)^2+u_0, \qquad v=\frac{1}{6}\ln(\tau-\tau_0)+v_0,
\end{equation}
where $\tau_0$, $u_0$, and $v_0$ are arbitrary constants. As a result we get
\begin{equation}
  a=a_0(\tau-\tau_0)^{(\Gamma+1)/(6\Gamma)}e^{3(\Gamma-1)(\tau-\tau_0)^2/\Gamma},\qquad \sigma=\sigma_0(\tau-\tau_0)^{-1/(6\Gamma)}e^{3(\tau-\tau_0)^2/\Gamma},
\end{equation}
where $a_0=e^{u_0+v_0}/\sigma_0 \ $, $\ \sigma_0=e^{(u_0-v_0)/\Gamma}$.

Another case with $k=0$ corresponds to $n=2+3\Gamma$. In this case
\begin{equation}
x=\frac{1}{6(\tau-\tau_0)},
\end{equation}
therefore,
\begin{equation}
u=\frac{1}{6}\ln(\tau-\tau_0)+u_0, \qquad v=3(\tau-\tau_0)^2+v_0,
\end{equation}
\begin{equation}
  a=a_0(\tau-\tau_0)^{(\Gamma-1)/(6\Gamma)}e^{3(\Gamma+1)(\tau-\tau_0)^2/\Gamma},\qquad \sigma=\sigma_0(\tau-\tau_0)^{1/(6\Gamma)}e^{-3(\tau-\tau_0)^2/\Gamma},
\end{equation}
where $a_0=e^{u_0+v_0}/\sigma_0$, $ \ \sigma_0=e^{(u_0-v_0)/\Gamma}$.

Let us note that $k=0$ corresponds to $\lambda=\pm 1$ and to the fourth case in Table~\ref{IntegrablePotentials} for~$\tilde{c}_2=0$.

\subsection{Solutions as functions of the cosmic time}

In the previous subsections we used the parametric time $\tau$ to linearize equations and get the general solutions for the induced gravity models with
power-law potentials. Let us note that
the  physically interesting question is the following one:  how will these solutions  look for an observer using the cosmic time?  In this subsection we show that some particular solutions have a simple form in the cosmic~time.

By definition the cosmic time $t$ is given~by
\begin{equation}
t = \int N(\tau) d\tau.
\label{time}
\end{equation}
So, $\tau$ is the cosmic time if and only if $N(\tau)\equiv 1$. The Hubble parameter is
\begin{equation*}
H(t)=\frac{d\ln(a)}{dt}.
\end{equation*}
The value of the Hubble parameter that corresponds to  constant solutions $\sigma_f$ at $n=2$ can be easily obtained
from (\ref{freq2}) with $N=1$ and $\ds=0$:
\begin{equation*}
H_{f}=\pm\frac{2\sqrt{\gamma c_0}}{\sqrt{3}}\sigma_{f}.
\end{equation*}

Let us consider solutions (\ref{sigma_a_x0}) that correspond to $x=x_0$ for other values of $n$. Using formula (\ref{H}), we obtain
\begin{equation}
\frac{\dot{N}}{N}=\frac{1-n}{\Gamma}\left(x_0-\frac{1}{x_0}\right).
\end{equation}
In the next section we consider solutions in cosmic time for the case $n=1$ in detail. For this reason in this subsection we consider the case $n\neq 1$ only. We obtain
\begin{equation}
N=N_0e^{(1-n)(x_0^2-1)\tau/(\Gamma x_0)},
\end{equation}
therefore,
\begin{equation}
t-t_0=\frac{\Gamma x_0 N_0}{(1-n)(x_0^2-1)} e^{(1-n)(x_0^2-1)\tau/(\Gamma x_0)}.
\end{equation}
We obtain that solution (\ref{sigma_a_x0}) becomes in the cosmic time:
\begin{equation}
 \sigma(t)=\left[\frac{(1-x_0^2)(n-1)}{N_0\Gamma x_0}(t-t_0)\right]^{\frac{1}{1-n}}=\left[\frac{\sqrt{(8\gamma n+10\gamma+3-2\gamma n^2)(6\gamma+1)}}{4\sqrt{\gamma^3c_0(n-1)^2(n-2)^2}(t-t_0)}\right]^{\frac{1}{n-1}}\!\!.
\label{sigma-power}
\end{equation}
The simplest way to find the corresponding $H(t)$ is to substitute the  $\sigma(t)$ obtained into the  system of equations (\ref{Fried1})--(\ref{Fried2}) with $N=1$. At $n\neq 2$ we obtain the following power-law solution
\begin{equation}
H(t)=\frac{2\gamma(1+n)+1}{2\gamma(n-1)(n-2)(t-t_0)}\,.
\label{Hubble-power}
\end{equation}

It is easy to check that the solution corresponding to the solution (\ref{sigma-power}), (\ref{Hubble-power}) in the minimally coupled model is the well-known particular solution \cite{exp-part}, i.e. the solution with $\phi \sim \ln t$ and $a \sim t^{p}$, where $p$ is some positive number.
 Indeed, on using the formulae from Section 2, one can see that the field $\phi$ behaves as a logarithm of the cosmic time $t$, while the lapse function $\tilde{N}$ behaves as a power of $t$. It then follows that the cosmic time $\tilde{t}$ is a function of $t$ and, hence, the cosmological radius $\tilde{a}$ is also a power-law function of $\tilde{t}$. Let us note that conditions on the conformal factor, which are necessary for the one-to-one correspondence between particular power-law solutions in the Einstein and Jordan frames, are given in~\cite{EPVZ}.

\section{An example: de Sitter solution in  induced gravity}

We have shown that if the exact general solution of  a cosmological model with a
minimally coupled scalar field and some potential $W(\phi)$ is known, then we can find the corresponding general solutions both in induced gravity models and in models with a conformally coupled scalar field plus the Hilbert--Einstein term, where the potential $V(\sigma)$ is given by the formula (\ref{poten1}).
In the preceding section we also have considered the induced gravity models with power-law self-interaction potential of the scalar field.
We would like to now explain in what sense  these solutions contain new physical information. It is here, that some differences between the frames can be~found.

It is well known that, in the model with a minimally coupled scalar field the de Sitter evolution can be realized if and only if the potential of the scalar field is a positive constant, or has an extremum for some value of the scalar field. The extremal value of the potential should also
be positive. Further, particular initial conditions on the scalar field should be imposed.
Namely, for the case of a constant potential it is enough to require a vanishing time derivative of the scalar field, while for the case of the potential with an extremum, it is necessary that the field itself has the value corresponding to this extremum.

For the case of the non-minimally coupled scalar field the situation is more complicated.
It is  known that if one considers a model with a non-minimally coupled scalar field with a  function  $U(\sigma)$ and with the self-interaction potential of the scalar field, proportional to this function squared $V(\sigma) = V_0U^2(\sigma)$, then, on requiring
$\dot{\sigma} = 0$, one obtains a de Sitter evolution, i.e. an exponential expansion.
This solution has, as its minimally coupled counterpart, the exponential evolution in the model with a constant potential. In both the models the scalar field has a constant value and one can easily check that the two solutions are connected by the relations described in Section 2.
As was shown in the preceding section this solution is an attractor for the class of general solutions, arising in the so called hyperbolic case for the induced gravity model.
Thus, the one-to-one relation between de Sitter solutions in the Jordan and Einstein frames is possible only if the corresponding scalar field $\sigma$ is a constant. Let us further note that for a solution with a constant $\sigma=\sigma_0$ in the Jordan frame (obtained in the case just discussed with  induced gravity and a quartic potential) the metric transformation (\ref{conf}) is trivial and the metric itself remains unchanged on choosing $U_0=U(\sigma_0)$. In such a case the Jordan
  frame is equivalent to the Einstein frame as far as homogeneous space-time evolution is concerned.

However, there are also other de Sitter solutions for the case of a non-minimal coupling between the scalar field and the curvature (see e.g. \cite{we-ind-ex,we-ind-ex1,Star-unpub,Sami:2012uh}).
For example, in the case of  induced gravity and of a self-interacting scalar field potential proportional to the  scalar field squared $V \sim \sigma^2$, a de Sitter solution with the scalar field changing in time exists.
As  is clear from  formula (\ref{exp-ind}) the counterpart of this potential is an exponential potential for the model with a minimally coupled scalar field. However, we know that the model with the exponential potential does not have a solution with a de Sitter regime.
Thus, we see that the family of cosmological evolutions in the minimally coupled model is physically different from the family of the corresponding evolutions in the  induced
 gravity model, obtained from the former by the  transformations, described in  Section~2.

 At this point an interesting question arises: what is the counterpart of the de Sitter solution obtained  in  the induced gravity model with the quadratic potential of self-interaction? Remarkably, this counterpart is nothing more than the well-known particular power-law solution of the equations of motion in the models with an exponential potential, which is expressible in terms of the cosmic time~\cite{exp-part}. To prove this statement we shall begin with the minimally coupled model
 with the exponential potential (\ref{pot-exp}).
We put $\tilde{N}=1$ in Eqs.~(\ref{Fried10}) and (\ref{Fried20}) and get the following equations
 
 \begin{equation}
 \tilde{H}^2 = \frac{\dot{\phi}^2}{3}+\frac23 W_0e^{2\sqrt{3}\lambda\phi},\qquad \dot{\tilde{H}} ={} -\dot{\phi}^2,
 \label{Fried4}
 \end{equation}
 where the Hubble parameter
 $\tilde{H}$  is defined with respect to the cosmic time $\tilde{t}=\tau$.

 The power-law solutions are
 \begin{equation}
 \label{power-law}
 \tilde{H}(\tilde{t}) = \frac{1}{3\lambda^2\tilde{t}},\qquad
   \phi(\tilde{t}) = {}-\frac{1}{\sqrt{3}\lambda}\ln\tilde{t}+\frac{1}{2\sqrt{3}\lambda}
   \ln\left(\frac{1-\lambda^2}{6W_0\lambda^4}\right).
  \end{equation}
We consider the model with $W_0>0$ and $0< \lambda^2 < 1$.
By a conformal transformation we can get from this model induced gravity models with power-law potentials of the form (\ref{exp-ind}). To get the induced gravity model with a quadratic potential
we  should choose $\lambda<0$ and
\begin{equation}
\gamma = \frac{3\lambda^2}{2(1-9\lambda^2)}.
\label{power-law2}
\end{equation}
A suitable model with a positive $\gamma$ exists for $-1/3<\lambda < 0$.
Using formula (\ref{connection1}), one~has
\begin{equation}
\sigma(\tilde{t}) =\sigma_1 \tilde{t},
\label{power-law3}
\end{equation}
where $\sigma_1$ is a constant.
Then
\begin{equation}
\sqrt{\frac{U_0}{U}} = \frac{1}{\sqrt{2\gamma}\sigma(\tilde{t})}=\frac{1}{\sqrt{2\gamma}
\sigma_1\tilde{t}}.
\label{power-law4}
\end{equation}
Hence, from Eq.~(\ref{Na}) it follows that in the induced gravity model  the lapse function and
the cosmological radius are
\begin{equation}
N = \frac{1}{\sqrt{2\gamma}
\sigma_1\tilde{t}},\qquad
a = \frac{\tilde{a}}{\sqrt{2\gamma}
\sigma_1\tilde{t}}.
\label{radius-power}
\end{equation}
We can now calculate the Hubble parameter value with respect to the cosmic time in the induced gravity model, which is given by the formula (\ref{time}):
\begin{equation}
\label{Ht}
H(t) = \frac{d\ln a}{dt} = \left(\tilde{H}-\frac{1}{\tilde{t}}\right)\frac{d\tilde{t}}{dt}
={}-\frac{(1-3\lambda^2)\sigma_1}{\lambda\sqrt{3(1-9\lambda^2)}}.
\end{equation}
Thereby, we see that the Hubble parameter in the induced gravity model is constant and hence, the evolution represents an exponential de Sitter expansion (naturally, the Hubble constant can also be negative and in this case one has a de Sitter contracting universe). Thus, we have shown that the de Sitter expansion in the induced gravity model with a quadratic self-interaction potential is transformed into the power-law expansion in the corresponding model with a minimally coupled scalar field  and with an exponential potential.

Generally, one can say (see e.g. \cite{Sasaki1} and an earlier work \cite{Dicke}) that
even if some classical phenomena look different in Einstein and Jordan frames, the relations
between different observables should be the same. Moreover, one can state \cite{Sasaki1}
that the choice between the Einstein and the Jordan frame is somewhat similar to the choice of the frame of reference in Newtonian mechanics. A coherent description of the laws of physics can be given in every frame of reference. Nonetheless a natural choice of such a frame of reference exists and is connected to each concrete physical problem. For example, a person on a merry-go round naturally describes physical phenomena around him, by taking into account inertial forces such as the centrifugal and that of Coriolis.
For this person the non-inertial rotating frame is the natural physical frame to be used. Analogously, in cosmology for a comoving observer the natural frame is that associated with his proper (i.e. cosmic) time.
Thus, since the transition between the Einstein and the Jordan frames implies the change of the cosmic time, the observed cosmological evolutions in these frames are different.

Finally we note that each scalar field potential in a minimally coupled model corresponds to a family of the induced gravity models with diverse self interacting potentials. For example the potential (see Table 1) $c_0e^{2\sqrt{3}\lambda\phi}$ generates the one parameter family of potentials $\tilde c_0 \sigma^{4+6\lambda \Gamma}$ having different exponents dependent on $\Gamma=\Gamma(\gamma)$. Hence, to the same cosmological evolution in a model with a minimally coupled scalar field corresponds a family of different evolutions in the corresponding induced gravity models.
For example, if $\gamma$ satisfies (\ref{power-law2}),
the the considering induced gravity model has exponential solution
(\ref{Ht}), otherwise it has power-law solutions
(\ref{Hubble-power}).
 The general formula connecting the cosmic time evolution of the Hubble parameter in the model with a minimally coupled scalar field to the cosmic time Hubble parameter evolutions in the associated induced gravity models is
\begin{equation}
H(t) = \sqrt{\frac{\gamma}{2U_0}}\sigma_0\exp\left(\frac{\phi}{\sqrt{12U_0}\Gamma}\right)
\times\left(\tilde{H}(\tilde{t})-\frac{\dot{\phi}}{\sqrt{12U_0}\Gamma\tilde{N}}\right).
\end{equation}

\section{Conclusions}
We have here shown that given the knowledge of  explicit general cosmological solutions for flat Friedmann models
with minimally coupled scalar fields one can construct the general solutions for  induced
gravity models and  models having a Hilbert--Einstein curvature term plus a scalar field conformally coupled to gravity. The latter are connected with the minimally coupled model by
the combination of a conformal transformation and a transformation of the scalar field.
The  forms of the self-interacting potentials for six explicitly integrable models
studied in~\cite{Fre} are presented here. We argue that although being mathematically in a one-to-one
correspondence  with the solutions in the minimally coupled models, the solutions
in the non-minimally coupled models are physically different, since the cosmological evolutions seen by an internal observer associated with the cosmic time can be quite different. We have given an explicit example of such a difference. Indeed, the de Sitter evolution in the induced gravity model has as its counterpart a power-law evolution in the corresponding model with a minimally coupled scalar field. This explains why a detailed study of cosmological evolutions in  exactly solvable models with a non-minimally coupled scalar field is interesting from the physical point of view.
We hope to undertake a detailed study of such integrable models in the close future \cite{future}.
We also hope to study the relation between the approach to the search of general cosmological solutions applied here and the approaches based on the explicit use of such symmetries as Noether symmetry \cite{Noether}, conformal  relations between different Lagrangians \cite{conf-Lagr} and the Hojman conservation quantities \cite{Hojman}.

\textbf{Acknowledgments. }  
Research of E.P. and S.V. is supported in part by the RFBR grant 14-01-00707 and by the Russian Ministry of Education and Science under grants NSh-3920.2012.2 and NSh-3042.2014.2, E.P. by the RFBR grant 12-02-31109. A.K. was partially supported by the RFBR grant 14-02-00894.


\begin{thebibliography}{72}
\bibitem{inflation}
Starobinsky A A 1986
Stochastic de sitter (inflationary) stage in the early universe 
{\it Lect. Notes Phys.} {\bf 246} 107--126 \\
Linde A D 1990 \textit{Particle Physics and Inflationary Cosmology} (Harwood, Chur, Switzerland)
\bibitem{dark}
Sahni V and Starobinsky A A 2000
The Case for a Positive Cosmological Lambda-term
{\it Int. J. Mod. Phys.} D  {\bf 9} 373 (arXiv: astro-ph/9904398)\\
Padmanabhan T 2003 Cosmological Constant - the Weight of the Vacuum
{\it Phys. Rep.} {\bf 380} 235 (arXiv:hep-th/0212290)\\
Piebles P J E  and Raptor B  2003 The Cosmological Constant and Dark Energy
{\it Rev. Mod. Phys.} {\bf 75} 559 (arXiv:astro-ph/0207347)\\
Sahni V 2002  	The Cosmological constant problem and quintessence {\it Class. Quantum Grav.} {\bf 19} 3435 (arXiv:astro-ph/0202076)\\
Copeland E J,  Sami M  and Tsujikawa S 2006 Dynamics of dark energy
{\it Int. J. Mod. Phys.} D {\bf 15} 1753 (arXiv:hep-th/0603057)\\
Sahni V and Starobinsky A A  2006 Reconstructing Dark Energy,
{\it Int. J. Mod. Phys. } D {\bf 15} 2105 (arXiv:astro-ph/0610026)\\
Tsujikawa S 2013 Quintessence: A Review
{\it Class. Quantum Grav.} {\bf 30} 214003 (arXiv:1304.1961[gr-qc])
\bibitem{cosmic}
Riess A \textit{et al.} 1998
 Observational evidence from supernovae for an accelerating universe and a cosmological constant.
{\it Astron. J.} {\bf 116}  1009 (arXiv: astro-ph/9805201)\\
Perlmutter S J  \textit{et al.} 1999
 Measurements of Omega and Lambda from 42 high redshift supernovae.
{\it Astrophys. J.} {\bf 517} 565 (arXiv: astro-ph/9812133)
\bibitem{gen-exp}
Salopek D S and Bond J R  1990	Nonlinear evolution of long wavelength metric fluctuations in inflationary models  {\it Phys. Rev. } D {\bf 42} 3936\\
Muller V, Schmidt H J and Starobinsky A A 1990 Power law inflation as an attractor solution for inhomogeneous cosmological models
 \textit{Class. Quantum Grav.} \textbf{7}  1163\\
Aguirregabiria J M  and  Chimento L P 1996 	Exact Bianchi type I models for an exponential-potential scalar field  {\it Class. Quantum Grav.} {\bf 13}  3197 \\
Chimento L P  1998 General solution to two-scalar field cosmologies with exponential potentials {\it Class. Quantum Grav.} {\bf 15}  965\\
Chimento L P, Cossarini A E and Zuccala N A 1998 Isotropic and anisotropic N-dimensional cosmologies with exponential potentials {\it Class. Quantum Grav.} {\bf 15}  57\\
Rubano C and Barrow J D 2001 Scaling solutions and reconstruction of scalar field potentials \textit{Phys. Rev. }  D \textbf{64}  127301 (arXiv:gr-qc/0105037)\\
Townsend P K 2003   Cosmic acceleration and M theory, arXiv:hep-th/0308149\\
Townsend P K and Wohlfarth M N R 2003
Accelerating cosmologies from compactification \textit{Phys. Rev. Lett.} \textbf{91}  061302 (arXiv:hep-th/0303097)\\
Emparan R and Garriga J 2003  A Note on accelerating cosmologies from compactifications and S branes	 	 \textit{J. High Energy Phys.} \textbf{0305}  028 (arXiv:hep-th/0304124)\\
Neupane I P 2004 Accelerating cosmologies from exponential potentials \textit{Class. Quantum Grav.} \textbf{21}  4383 (arXiv:hep-th/0311071)\\
Russo J G 2004 Exact solution of scalar tensor cosmology with exponential potentials and transient acceleration \textit{Phys. Lett. } B  \textbf{600} 185 (arXiv:hep-th/0403010)\\
Elizalde E, Nojiri S and Odintsov S D 2004 Late-time cosmology in (phantom) scalar-tensor theory: Dark energy and the cosmic speed-up \textit{Phys. Rev. } D \textbf{70}  043539 (arXiv:hep-th/0405034)\\
Chimento L P 2004 Extended tachyon field, Chaplygin gas and solvable k-essence cosmologies \textit{Phys. Rev.} D \textbf{69} (2004) 123517 (arXiv:astro-ph/0311613)\\
Chimento L P and Forte M I  2006 Anisotropic k-essence cosmologies \textit{Phys. Rev.} D \textbf{73}  063502 (arXiv:astro-ph/0510726)\\
Chimento L P, Forte M I, Lazkoz R and Richarte M G 2009 Internal space structure generalization of the quintom cosmological scenario
\textit{Phys. Rev.} D \textbf{79}  043502 (arXiv:0811.3643[astro-ph])\\
Dudas E, Kitazawa N and Sagnotti A 2010 On Climbing Scalars in String Theory \textit{Phys. Lett.} B \textbf{694} (2010) 80 (arXiv:1009.0874[hep-th])\\
Andrianov A A, Cannata F and Kamenshchik A Yu 2011 General solution of scalar field cosmology with a (piecewise) exponential potential \textit{J. Cosmol. Astropart. Phys.} \textbf{1110} (2011) 004 (arXiv:1105.4515[gr-qc])\\
Chimento L P, Lazkoz R and Richarte M G 2011 Enhanced Inflation in the Dirac-Born-Infeld framework, \textit{Phys. Rev.} D \textbf{83} 063505 (arXiv:1011.2345[astro-ph.CO])\\
Dudas E, Kitazawa N,  Patil S P and Sagnotti A 2012 CMB Imprints of a Pre-Inflationary Climbing Phase \textit{J. Cosmol. Astropart. Phys.} \textbf{1205}  012 (arXiv:1202.6630[hep-th])\\
Andrianov A A,  Cannata F  and Kamenshchik A Yu 2012 Remarks on the general solution for the flat Friedman universe with exponential scalar-field potential and dust  \textit{Phys. Rev. D} \textbf{86}  107303 (arrive:1206.2828[gr-qc])


\bibitem{exp-part}
Lucchin F and Matarrese S 1985 Power Law Inflation \textit{Phys. Rev.} D \textbf{32}  1316\\
Halliwell J J 1987 Scalar Fields in Cosmology with an Exponential Potential \textit{Phys. Lett.} B \textbf{185}  341\\
 Barrow J D  1987 Cosmic No Hair Theorems and Inflation \textit{Phys. Lett.} B \textbf{187} 12\\
Burd A B  and Barrow J D 1988 Inflationary Models with Exponential Potentials \textit{Nucl. Phys.} B \textbf{308}  929\\
Ellis G F R and Madsen M S 1991 Exact scalar field cosmologies \textit{Class. Quantum Grav.} \textbf{8}  667\\
Gorini V, Kamenshchik A Yu, Moschella U and Pasquier V 2004 Tachyons, scalar fields and cosmology
\textit{Phys. Rev.} D \textbf{69} 123512 (arXiv:hep-th/0311111)


\bibitem{hyper}
de~Ritis R, Marmo G, Platania G, Rubano C, Scudellaro P  and Stornaiolo C 1990
New approach to find exact solutions for cosmological models with a scalar field \textit{Phys. Rev.} D \textbf{42}  1091\\
Capozziello S, de~Ritis R, Rubano C and Scudellaro P 1996 \textit{Riv. Nuovo Cim.} \textbf{19}  1\\
Piedipalumbo E, Scudellaro P, Esposito G and Rubano C 2012 On quintessential cosmological models and exponential potentials \textit{Gen. Relat. Grav.} \textbf{44}  2611 (arXiv:1112.0502[astro-ph.CO])
\bibitem{Fre}
 Fr\'e P, Sagnotti A and Sorin A S 2013
  Integrable Scalar Cosmologies I. Foundations and links with String Theory,
\textit{Nucl. Phys.} B \textbf{877}  1028 (arXiv:1307.1910[hep-th])
\bibitem{Fre2} Fr\'e P, Sorin A S and  Trigiante M 2014
Integrable Scalar Cosmologies II. Can they fit into Gauged Extended Supergavity or be encoded in N=1 superpotentials?, \textit{Nucl. Phys.} B  \textbf{881}  91--180 (arXiv:1310.5340 [hep-th])


\bibitem{Onefield}
Muslimov A G 1990
On the Scalar Field Dynamics in a Spatially Flat Friedman Universe,
\textit{Class. Quantum Grav.} \textbf{7}  231 \\
Zhuravlev V M, Chervon S V  and Shchigolev V K 1998
New classes of exact solutions in inflationary cosmology,
  \textit{J.\ Exp.\ Theor.\ Phys.}   \textbf{87}  223\\
Aref'eva I Ya, Koshelev A S and Vernov S Yu 2006
Exact solution in a string cosmological model
\textit{Theor. Math. Phys.} \textbf{148}  895 (arXiv:astro-ph/0412619)\\
Guo Z K , Ohta N  and Zhang Y Z 2005
Parametrization of quintessence and its potential,
\textit{Phys.\ Rev.}  D \textbf{72}  023504 (arXiv:astro-ph/0505253)\\
Capozziello S, Nojiri S and Odintsov S D 2006
Dark Energy: the equation of state description versus scalar-tensor or modified gravity
\textit{Phys. Lett.} B \textbf{634}  93 (arXiv:hep-th/0512118)\\
Bazeia D, Gomes C B, Losano L  and Menezes R 2006
First-order formalism and dark energy,
\textit{Phys. Lett.} B \textbf{633}  415 (arXiv:astro-ph/0512197)\\
Townsend P K 2008
Hamilton-Jacobi Mechanics from Pseudo-Supersymmetry
\textit{Class. Quantum Grav.} 25  045017 (arXiv:0710.5178[hep-th])\\
Yurov A V, Yurov V A, Chervon S V and  Sami M 2011
Total energy potential as a superpotential in integrable cosmological models,
\textit{Theor. Math. Phys.} 166  259\\
Kamenshchik A Yu and Manti S 2012
Scalar field potentials for closed and open cosmological models
\textit{Gen. Relat. Grav.} \textbf{44}  2205 (arXiv:1111.5183[gr-qc])\\
Kim H-Ch 2013
Exact solutions in Einstein cosmology with a scalar field
\textit{Mod. Phys. Lett.} A \textbf{28}  1350089 (arxiv:1211.0604[gr-qc])\\
Harko T, Lobo F S N and Mak M K 2014 Arbitrary scalar field and quintessence cosmological models {\it Eur. Phys. J.} C {\bf 74} 2784 (arXiv:1310.7167[gr-qc])

\bibitem{Two-fields}
Aref'eva I Ya, Koshelev A S and Vernov S Yu 2005
Crossing the $w=-1$ barrier in the D3-brane dark energy model
{\it Phys. Rev. D} {\bf 72}  064017  (arXiv:astro-ph/0507067)\\
Vernov S Yu 2008
Construction of exact solutions in two-field cosmological models
 {\it Theor. Math. Phys.} {\bf 155}  544  (arXiv:astro-ph/0612487)\\
Andrianov A A, Cannata F, Kamenshchik A Yu and Regoli D 2008
Reconstruction of scalar potentials in two-field cosmological models,
{\it J. Cosmol. Astropart. Phys.} {\bf 0802}  015 (arXiv:0711.4300[gr-qc])\\
Aref'eva I Ya, Bulatov N V and Vernov S Yu 2010
 Stable exact solutions in cosmological models with two scalar fields
{\it Theor. Math. Phys.} {\bf 163}  788 (arXiv:0911.5105[hep-th])

\bibitem{we-ind-ex}
Kamenshchik A Yu, Tronconi A and Venturi G 2011
 Reconstruction of scalar potentials in induced gravity and cosmology
{\it Phys. Lett. B} {\bf 702}  191 (arXiv:1104.2125[gr-qc])
\bibitem{we-ind-ex1}
Kamenshchik A Yu, Tronconi A, Venturi G and Vernov S Yu 2013
 Reconstruction of Scalar Potentials in Modified Gravity Models
{\it Phys. Rev. D} {\bf 87}  063503 (arXiv:1211.6272[gr-qc])

\bibitem{Sakharov}
 Sakharov A D 1967 {\it Dok. Akad. Nauk SSSR} {\bf 117}  70, {\it Sov. Phys. Dokl.} {\bf 12}  1040
\bibitem{induced}
Cooper F and Venturi G  1981 Cosmology and broken scale invariance
{\it Phys. Rev. } D {\bf 24}  3338\\
Finelli F, Tronconi A and Venturi G  2008
Dark Energy, Induced Gravity and Broken Scale Invariance
{\it Phys. Lett. } B  {\bf 659}  466  (arXiv:0710.2741[astro-ph])\\
Cerioni A, Finelli F, Tronconi A and Venturi G 2009
Inflation and Reheating in Induced Gravity
{\it Phys. Lett. } B {\bf 681} 383 (arXiv:0906.1902[astro-ph.CO])\\
Cerioni A, Finelli F, Tronconi A and Venturi G 2010
 Inflation and Reheating in Spontaneously Generated Gravity
{\it Phys. Rev. } D {\bf 81}  123505  (arXiv:1005.0935[gr-qc])\\
Tronconi A and Venturi G 2011
Quantum Back-Reaction in Scale Invariant Induced Gravity Inflation
{\it Phys. Rev. } D {\bf 84}  063517 (arXiv:1011.3958[gr-qc])\\
 Kamenshchik A Yu, Tronconi A and Venturi G 2012
 Dynamical Dark Energy and Spontaneously Generated Gravity
{\it Phys. Lett. } B {\bf 713}  358  (arXiv:1204.2625[gr-qc])
\bibitem{nonmin-inf}
Spokoiny B L 1984
Inflation and Generation of Perturbations in Broken Symmetric Theory of Gravity
{\it Phys. Lett.} {\bf 147 B}  39\\
Futamase T and Maeda K-i 1989
Chaotic Inflationary Scenario in Models Having Nonminimal Coupling with Curvature
{\it Phys. Rev. } D {\bf 39}  399\\
Salopek D S, Bond J R and  Bardeen J M 1989
Designing Density Fluctuation Spectra in Inflation
 {\it Phys. Rev. } D {\bf 40} 1753\\
Fakir R and Unruh W G 1990
Improvement on cosmological chaotic inflation through nonminimal coupling
{\it Phys. Rev. } D {\bf 41} 1783

\bibitem{nonmin-quant}
Barvinsky A O and Kamenshchik A Yu 1994 Quantum scale of inflation and particle physics of the early universe
{\it Phys. Lett. } B {\bf 332}  270  (arXiv:gr-qc/9404062)\\
Barvinsky A O, Kamenshchik A Yu, Kiefer C and Steinwachs C F 2010 Tunneling cosmological state revisited: Origin of inflation with a non-minimally coupled Standard Model Higgs inflaton
 {\it Phys. Rev.} D {\bf 81} 043530 (arXiv:0911.1408[hep-th]).


\bibitem{Higgs}
Bezrukov F L  and Shaposhnikov M 2008
The Standard Model Higgs boson as the inflaton 
{\it Phys. Lett. B} {\bf 659} 703 (arXiv:0710.3755[hep-th])\\
Barvinsky A O, Kamenshchik A Yu and Starobinsky A A 2008
Inflation scenario via the Standard Model Higgs boson and LHC
{\it J. Cosmol. Astropart. Phys.} {\bf 0811} 021 (arXiv:0809.2104[hep-ph])\\
Bezrukov F L, Magnin A and Shaposhnikov M 2009
Standard Model Higgs boson mass from inflation
{\it Phys. Lett. B} {\bf 675} 88 (arXiv:0812.4950[hep-ph])\\
De~Simone A, Hertzberg M P and Wilczek F 2009
Running Inflation in the Standard Model
{\it Phys. Lett. } B {\bf 678} 1 (arXiv:0812.4946[hep-ph])\\
Barvinsky A O, Kamenshchik A Yu, Kiefer C, Starobinsky A A and Steinwachs C
2009 Asymptotic freedom in inflationary cosmology with a non-minimally coupled Higgs field
{\it J. Cosmol. Astropart. Phys.} {\bf 0912} 003 (arXiv:0904.1698[hep-ph])\\
Bezrukov F L, Magnin A, Shaposhnikov M and Sibiryakov S
Higgs inflation: consistency and generalizations
2011 {\it J. High Energy Phys.} {\bf 1101}  016 (arXiv:1008.5157[hep-ph])\\
Barvinsky A O, Kamenshchik A Yu, Kiefer C, Starobinsky A A and Steinwachs C F 2012
Higgs boson, renormalization group, and naturalness in cosmology
{\it Eur. Phys. J. C} {\bf 72}  2219  (arXiv:0910.1041[hep-ph])\\
Bezrukov F and Gorbunov D 2013 Light inflaton after LHC8 and WMAP9 results
{\it J. High Energy Phys.} {\bf 1307} 140  (arXiv:1303.4395[hep-ph])\\
Bezrukov F  2013 The Higgs field as an inflaton  {\it Class. Quant. Grav.} {\bf 30}  214001 (arXiv:1307.0708[hep-ph])

\bibitem{discovery}
Aad G et al., [ATLAS Collaboration] 2012 Observation of a new particle in the search for the Standard
Model Higgs boson with the ATLAS detector at the LHC {\it Phys. Lett. B} {\bf 716} 1  (arXiv:1207.7214[hep-ex])\\
Chatrchyan S et al., [CMS Collaboration] 2012 Observation of a new boson at a mass of 125 GeV with the CMS
experiment at the LHC {\it Phys. Lett. B} {\bf 716} 30 (arXiv:1207.7235[hep-ex])


\bibitem{Wagoner}
Wagoner R W 1970 Scalar--Tensor Theory and Gravitational Waves {\it Phys. Rev.} D {\bf 1}  3209
\bibitem{Maeda}
Maeda K i 1989  	
Towards the Einstein--Hilbert Action via Conformal Transformation
{\it Phys. Rev.} D {\bf 39}  3159
\bibitem{Bronnikov}
 Bronnikov K A 1973
  Scalar--tensor theory and scalar charge
  {\it Acta Phys. Polon.} B {\bf 4}  251


\bibitem{Dicke}
Dicke R H 1962 Mach's Principle and Invariance under Transformation of Units 
{\it Phys. Rev.} {\bf~125}~2163

\bibitem{Sasaki}
Deruelle N, Sasaki M and Sendouda Y 2008
  'Detuned' f(R) gravity and dark energy
  {\it Phys.\ Rev.\ D} {\bf 77} 124024
  (arXiv:0803.2742[gr-qc])
\bibitem{Sasaki1}
Deruelle N and Sasaki M 2011
  Conformal equivalence in classical gravity: the example of 'Veiled' General Relativity
  {\it Springer Proc.\ Phys.}  {\bf 137}  247 (arXiv:1007.3563[gr-qc])
\bibitem{frames}
Faraoni V and  Nadeau Sh 2007
(Pseudo)issue of the conformal frame revisited
\textit{Phys. Rev. } D \textbf{75}  023501 (arXiv:gr-qc/0612075)\\
Capozziello S, Nojiri Sh, Odintsov S D and Troisi A 2006
Cosmological viability of f(R)-gravity as an ideal fluid and
its compatibility with a matter dominated phase
\textit{Phys.\ Lett. } B \textbf{639}  135 (arXiv:astro-ph/0604431)\\
Lerner R N and McDonald J 2010
Higgs Inflation and Naturalness
{\it J. Cosmol. Astropart. Phys.} {\bf 1004} 015  (arXiv:0912.5463[hep-ph])\\
Hertzberg M P 2010
On Inflation with Non-minimal Coupling
{\it J. High Energy Phys.} {\bf 1011} 023   (arXiv:1002.2995[hep-ph])\\
Kaiser D I 2010
Conformal Transformations with Multiple Scalar Fields
{\it Phys. Rev. D} {\bf 81} 084044  (arXiv:1003.1159[gr-qc])\\
Steinwachs C F and Kamenshchik A Yu 2011
One-loop divergences for gravity non-minimally coupled to a multiplet of scalar fields: calculation in the Jordan frame. I. The main results
{\it Phys. Rev. } D {\bf 84}  024026  (arXiv:1101.5047[hep-th])\\
Steinwachs C F and Kamenshchik A Yu 2012
Non-minimal Higgs Inflation and Frame Dependence in Cosmology
 {\it AIP Conf. Proc.} {\bf 1514} 161 (arXiv:1301.5543[hep-th])

\bibitem{Faraoni}
Faraoni V, Gunzig E and Nardone P 1999
 Conformal transformations in classical gravitational theories and in cosmology
  {\it Fund. Cosmic Phys.}  {\bf 20}  121 (arXiv:gr-qc/9811047)
\bibitem{Bron-Meln}
Bronnikov K A and Melnikov V N 2004
         	Conformal frames and D-dimensional gravity
        in: Proceedings of the 18th Course of the School on Cosmology and
        Gravitation: The Gravitational Constant, Generalized Gravitational
        Theories and Experiments (30 April-10 May 2003, Erice).
        Ed. G.T. Gillies, V.N. Melnikov and V. de Sabbata,
        Kluwer, Dordrecht/Boston/London,  pp. 39--64
        (arXiv:gr-qc/0310112)


\bibitem{Star-unpub}
Starobinsky A A, in the Proceedings of the Cosmology Workshop Montpellier 12 (unpublished).
\bibitem{KKhT} Kamenshchik A Yu, Khalatnikov I M and  Toporensky A V 1997
Complex inflaton field in quantum cosmology
{\it Int. J. Mod. Phys. } D {\bf 6} 649 (arXiv:gr-qc/9801039)
\bibitem{ABGV}
Aref'eva I Ya, Bulatov N V, Gorbachev R V and  Vernov S Yu 2014
Non-minimally Coupled Cosmological Models with the Higgs-like Potentials and Negative Cosmological Constant,
\textit{Class. Quantum Grav.} \textbf{31} 065007 (arXiv:1206.2801[gr-qc])
\bibitem{BNOS}
Bamba K, Nojiri Sh, Odintsov S D  and S\'aez-G\'omez D 2014
Possible antigravity regions in $F(R)$ theory? \textit{Physics Letters} B \textbf{730} 136--140  (arXiv:1401.1328[hep-th])
\bibitem{EPVZ}
Elizalde E, Pozdeeva E O, Vernov S.Yu. and  Zhang Y.-li 2013
Cosmological Solutions of a Nonlocal Model with a Perfect Fluid
{\it J. Cosmol. Astropart. Phys.} {\bf 1307}  034  (arXiv:1302.4330[hep-th])
\bibitem{Sami:2012uh}
Sami M, Shahalam M, Skugoreva M and  Toporensky A 2012
Cosmological dynamics of non-minimally coupled scalar field system and its late time
cosmic relevance
{\it Phys. Rev. } D {\bf 86} 103532  (arXiv:1207.6691[hep-th])
\bibitem{future}
Kamenshchik A Yu, Pozdeeva E O, Tronconi A, Venturi G and Vernov S Yu, work in progress.
\bibitem{Noether}
Paliathanasis A, Tsamparlis M, Basilakos S and Capozziello S 2014
  Scalar-Tensor Gravity Cosmology: Noether symmetries and analytical solutions
  (arXiv:1403.0332[astro-ph.CO])
\bibitem{conf-Lagr}
Tsamparlis M, Paliathanasis A, Basilakos S and Capozziello S 2013
  Conformally related metrics and Lagrangians and their physical interpretation in cosmology
  {\it Gen. Rel. Grav.}  {\bf 45}  2003
  (arXiv:1307.6694[gr-qc])
\bibitem{Hojman}
Capozziello S and Roshan M 2013
 Exact cosmological solutions from Hojman conservation quantities
 {\it  Phys. Lett.} B {\bf 726}  471 (arXiv:1308.3910[gr-qc])
\end{thebibliography}
\end{document}